\documentclass[12pt, a4paper]{article}
\usepackage{amsmath}
\usepackage{amsthm}
\usepackage{amsfonts}
\usepackage{amssymb,amsmath,latexsym}
\usepackage{graphicx}
\input epsf 

\newtheorem{theorem}{Theorem}[section]

\theoremstyle{definition}

\def\C{{\mathbb C}}  

\def\R{{\mathbb R}}    
\def\supp{\operatorname{supp}}  
\begin{document}
\begin{titlepage}
\vspace*{7mm}

\begin{center}
{\bf \Large The Correlator Toolbox, Metrics and Moduli} \\
\vspace*{8mm}

{Robert~O. Bauer$^\mathrm{a}$  and   
Roland~M. Friedrich$^\mathrm{b}$ 
} \\

\vspace*{3mm}

${}^\mathrm{a}$
{\em Department of Mathematics\\ University of Illinois at Urbana-Champaign\\ 1409 West Green Street \\ Urbana, IL 61801, USA} \\
\vspace{2mm}
${}^\mathrm{b}$
{\em Max-Planck-Institut f\"ur Mathematik, D-53111 Bonn} \\
\vspace*{6mm}

\end{center}

\begin{abstract}
We discuss the possible set of operators from various boundary conformal field theories to build meaningful correlators that lead via a Loewner type procedure to generalisations of SLE($\kappa,\rho$). We also highlight the necessity of moduli for a consistent kinematic description of these more general stochastic processes.  As an illustration  
we give a geometric derivation of $\text{SLE}(\kappa,\rho)$ in terms of
conformally invariant random growing compact subsets of polygons. The
parameters $\rho_j$ are related to the exterior angles of the polygons. We
also show that $\text{SLE}(\kappa,\rho)$ can be generated by a Brownian
motion in a gravitational background, where the metric and the Brownian motion are coupled. The metric is obtained as the 
pull-back of the Euclidean metric of a fluctuating polygon.      

\vfill

\begin{tabular}{ll}
{\em PACS 2003:} &
02.50.Ey, 
05.50.+q, 
11.25.Hf  
\\
{\em MSC 2000:}   &
60D05, 
58J65, 
81T40 
\\
{\em Keywords:}  & Probability Theory; Conformal Field Theory \\
 & \\
{\em Email:}    &  {\tt rbauer@math.uiuc.edu, rolandf@mpim-bonn.mpg.de}
\end{tabular}
\end{abstract}
\end{titlepage}

\section{Initial Considerations}
The similarity of expressions from $\text{SLE}(\kappa,\rho)$~\cite{LSW:2003,dubedat:2003}  with correlators in the Coulomb gas formalism was (probably) first noticed by S.~Chakravarty~\cite{ChS2004}. His questions for an explanation of this ``coincidence" led to some (almost) unpublished notes~\cite{RF:2004}.  
An independent  explanation of $\text{SLE}(\kappa,\rho)$ from the CFT perspective was given in~\cite{Cardy2004}.

In this paper we will show, thereby relying on the results in~\cite{BF:2005b},  that $\text{SLE}(\kappa,\rho)$ arises naturally
when one considers random growing compacts in polygons, and how this fits into the context of conformal field theories.  We will also outline, that there are more meaningful stochastic processes of SLE($\kappa,\rho$) type, that can be derived from physical considerations. 

For the mathematical details concerning the ``fluctuating polygons",  see~\cite{BF:2005b}. For the relations of SLE to diffusion processes on moduli spaces and / or general CFT, see~\cite{BF:2004a, BF:2005a,Cardy:2005, FK, KBonn}.
\subsection{SLE($\kappa, \rho$)}
$\text{SLE}(\kappa,\rho)$ was introduced on mathematical grounds, as a generalisation of  ``ordinary" SLE,  in the landmark paper~\cite{LSW:2003} by Lawler, Schramm and Werner, and further studied in \cite{werner:2004,dubedat:2003}. (Extensive mathematical details on SLE can be found, e.g. in~\cite{lawlerbook}).

Stochastic Loewner evolution (or SLE) as introduced by Schramm in 
\cite{schramm:2000} describes random growing compacts, in simply connected 
planar domains, which correspond (supposedly) to the conformally invariant scaling limit of discrete random simple
curves that also satisfy a Markovian-type property. Then by the two above properties (plus a
reflection symmetry) SLE is canonical in the sense
that there exists only a one-parameter family of random non-selfcrossing
curves $\gamma$ with these properties. The dynamical way to describe the measures, is by solving
L\"owner's equation \cite{loewner:1923} with a driving function given in terms
of Brownian motion. 
So for the upper half-plane
$\mathbb H$, and $\kappa\ge0$, consider for each $z\in\overline{\mathbb H}$ the ordinary
differential equation
\begin{equation}\label{E:CSLE}
\partial_t g_t(z)=\frac{2}{g_t(z)-W_t}, \quad g_0(z)=z,
\end{equation}
where $W_t=\sqrt{\kappa}B_t$, and $B_t$ is a 
one-dimensional standard Brownian motion. Let $T_z$ be the duration for which this
equation is well defined, i.e. $T_z=\sup\{t:\inf_{s\in[0,t]}|g_t(z)-W_t|>0\}$,
and set $K_t=\{z:T_z\le t\}$. Then one can show that $g_t$ is a conformal
map from $\mathbb H\backslash K_t$ onto $\mathbb H$ with
$\lim_{z\to\infty}(g(z)-z)=0$. It can also be shown \cite{RS} that with
probability one the random
growing compact set $K_t$ is generated by a random non-selfcrossing curve
$t\mapsto\gamma_t$ in the sense that $\mathbb H\backslash K_t$ is the unbounded
component of $\mathbb H\backslash\gamma[0,t]$. $\gamma$ is a random curve
connecting the boundary points $0$ and $\infty$ and is called {\em chordal}
$\text{SLE}_{\kappa}$ in $\mathbb H$ from $0$ to $\infty$. 

For calculations involving SLE conformal invariance is a powerfull tool as it is always permissible to choose the geometrically most convenient
configuration to do a given calculation, where the solution depends only on the conformal equivalence class, or the moduli, of the configuration.

Now, let $z_1< z_2<\dots< z_n$ be real numbers, all distinct
from $0$. Consider the system of stochastic differential equations
\begin{align}\label{E:rho-diff}
        dW_t&=\sqrt{\kappa}\ dB_t
        +\sum_{k=1}^n\frac{\rho_k}{W_t-Z_t^k}\ dt\notag\\
        dZ_t^k&=\frac{2}{Z_t^k-W_t}\ dt,\quad k=1,\dots, n~,
\end{align}
with $W_0=0, Z_0^1=z_1,\dots, Z_0^n=z_n$, and where $B_t$ is a one-dimensional
\begin{figure}[ht]
\begin{center}
\includegraphics[scale=0.5]{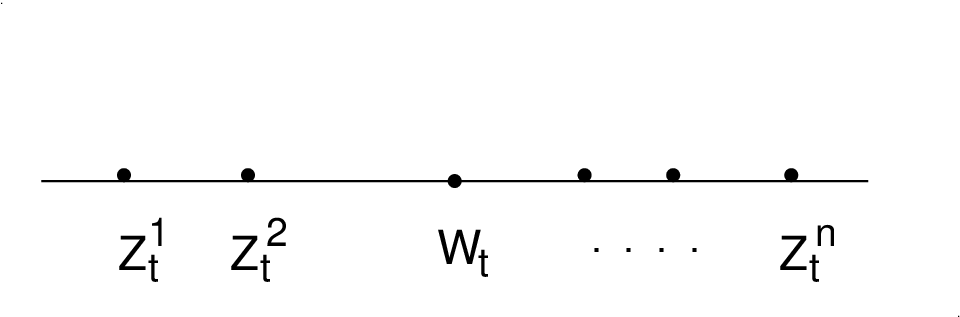}
\caption{The SLE($\kappa,\rho$) random dynamical system on $\R$.}
\label{c2}
\end{center}
\end{figure}
standard Brownian motion. Then, at least up to some small time $t$, the solution will exist. As above, let $g_t(z)$ be the solution to \eqref{E:CSLE}. Then the family
of conformal maps $g_t$ is called $\text{SLE}(\kappa,\rho)$ in the upper
half-plane from 
$(0,z_1,\dots, z_n)$ to $\infty$.

\subsection{From Physics}
Let us start with aspects of Liouville field theory~\cite{Pol81}, (for an early review~\cite{GM}), that is related to the problem of quantum gravity. There, one of the major tasks is to properly integrate over all metrics modulo diffeomorphisms.   
  
So, let us consider a two-dimensional surface $M$, possibly with boundary $\partial M$ and a Riemannian metric $g$, i.e. a bordered Riemann surface. 
In the conformal gauge
any metric can be written as 
\begin{equation}
\label{S1.1}
g=e^{\gamma\phi} g_0~,
\end{equation}
where $\gamma$ is a parameter and $g_0$ is the ``reference metric", which also determines a point $[g_0]$ in the moduli space. The field $\phi$ is known as the Liouville mode.

For a closed surface $M$ the underlying field theory  is given by a  bulk action
\begin{equation}
\label{S1.2}
S[g_0,\phi]:=\frac{1}{8\pi}\int_M \sqrt{g_0}\left((\nabla\phi)^2+Q \phi R(g_0)+\frac{\mu}{\gamma^2} e^{\gamma\phi}\right)\, d^2z~,
\end{equation} 
with the coupling constant $\gamma (\hbar=\gamma)$, $R(\cdot)$ the scalar curvature of $(M,g_0)$ (in dimension two $R$ is twice the sectional curvature) and the cosmological constant $\mu>0$ .

Let us point out, that in the above action~(\ref{S1.2}), the $Q$-term represents the action in the Coulomb gas formalism (CGF) on closed Riemann surfaces, i.e. a Gaussian conformal field theory in the presence of an imaginary background charge with $Q=2i\alpha_0$. Its presence leads to a modified stress-energy tensor
\begin{equation}
\label{M1.15}
T=-\frac{1}{2}\partial\phi\,\partial\phi+i\alpha_0\,\partial^2\phi~,
\end{equation}
which generates a Virasoro algebra with central charge
\begin{displaymath}
c=1-12\alpha^2_0~.
\end{displaymath}

Further, let us mention  that~(\ref{S1.2}) also represents a quantum conformal field theory. But contrary to ``standard" CFT where we have the state-field correspondence, this does not hold any longer in Liouville field theory. Here the primary operators are of the form $e^{\alpha\phi}$ with conformal weight $-\frac{1}{2}(\alpha-\frac{Q}{2})^2+\frac{Q^2}{8}$, and the set of operators and the set of states are distinct.

Now let us come to the case of surfaces with boundaries. There are two major new aspects. First the bulk action has to be extended by contributions from the boundary and second, one has to fix boundary conditions.

So, let us define the boundary action as 
\begin{equation}
\label{M3.82}
S_L\rightarrow S_{\text{bulk}}+\frac{Q}{8\pi}\int_{\partial M}\phi k\, |d z|+\frac{\lambda}{4\pi\gamma^2}\int_{\partial M} e^{\frac{1}{2}\gamma\phi}\, |d z|~,
\end{equation}
where $k$ is the geodesic curvature of the boundary, $|d z|$ the line element and $\lambda$ the boundary cosmological constant. 

To have a well posed variational problem, the possible boundary conditions are Dirichlet or Neumann, or combinations of the two. 

Then the bulk equation of motion $\frac{\delta S}{\delta\phi}$ is:
\begin{eqnarray}
\label{S2.2}
R(g) & = &  -\frac{\mu}{2}~,\qquad\text{which is equivalent to}\\\nonumber
\Delta\gamma\phi & = & \frac{\mu}{2}\, e^{\gamma\phi}+R(g_0)~,
\end{eqnarray}
i.e. the metric $g=e^{\gamma\phi}g_0$ has constant negative curvature. The stress-energy tensor is found by varying the action with respect to the reference metric, i.e. $T_{ab}=2\pi\frac{\delta S}{\delta g_0^{ab}}$, which results in
\begin{eqnarray}\label{M3.7}
T_{z\bar{z}} & = & 0~, \\\nonumber
T_{zz} & = & -\frac{1}{2}(\partial\phi)^2+\frac{1}{2}Q\,\partial^2\phi~.
\end{eqnarray}
As $\phi$ is a component of a metric as well, it transforms under conformal mappings $z\mapsto w=f(z)$ like
\begin{equation}
\label{M3.8}
\phi\mapsto\phi+\frac{1}{\gamma}\log\left|\frac{dw}{dz}\right|^2~.
\end{equation}
In particular, the $U(1)$ current $\partial_z\phi$ transforms as
\begin{equation}
\label{M3.9}
\partial_z\phi\mapsto\frac{dw}{dz}\partial_w\phi+\frac{d}{dz}\frac{1}{\gamma}\log\left|\frac{dw}{dz}\right|~,
\end{equation}
and the stress tensor $T_{zz}$ as
\begin{equation}
\label{M3.10}
T_{zz}\mapsto\left(\frac{dw}{dz}\right)^2 T_{ww}+\frac{1}{\gamma^2}\{w;z\}~.
\end{equation}
Here, $\{w;z\}$ denotes the Schwarzian derivative. 

The equations of motion in the bordered case, with ($\delta\phi|\partial M=0$) are:
\begin{equation}
\label{M3.38}
\frac{\partial(\gamma\phi)}{\partial{\bf n}}+k+\frac{\lambda}{2} e^{\frac{1}{2}\gamma\phi}=0~,
\end{equation}
where the {\bf n}-term denotes the normal derivative. We note, that like 
in the case of the bulk, there are also vertex operators on the boundary with some conformal weight $\Delta_{\partial M}=-2\alpha^2+Q\alpha$. 

In the above discussions we tacitly assumed, that the metric has no singularities neither in the bulk nor on the boundary. We shall shortly see, how that changes things. 

\subsection{The Uniformisation Problem}
The solutions of the classical equations of motion~(\ref{S2.2}) and~(\ref{M3.38}), i.e. the Liouville equation, are intrinsically related to the uniformisation problem of Riemann surfaces. It states, that every Riemann surface is conformally equivalent to either the Riemann sphere, the upper half-plane $\mathbb{H}$ or to a quotient of $\mathbb{H}$ by some discrete subgroup $\Gamma\subset SL(2,\R)$.

The ``fundamental" solution of the Liouville equation for $\mathbb{H}$ is the Poincar\'e metric with constant negative curvature $-1$. The classical solutions in Euclidean space of the Liouville equations are of the form
\begin{equation}
\label{S2.12}
e^{\gamma\phi}|dz|^2=\frac{4}{\mu}\frac{\partial A\,\bar{\partial} B}{\left(A(z)-B(\bar{z})\right)^2}\,|dz|^2
\end{equation}
with $A$ and $B$ some (locally) defined functions of $z,\bar{z}$. 

If $X=\mathbb{H}/\Gamma$ denotes the quotient by a discrete subgroup, then there exists a natural projection $\pi:\mathbb{H}\rightarrow X$ with an ``inverse" map
\begin{equation}
\label{M3.15}
f:X\rightarrow\mathbb{H}~,
\end{equation}
depending on the moduli of $X$. In terms of the inverse map $f$ the solution of the field equation~(\ref{S2.12}) has energy-momentum tensor (Fuchsian projective connection) satisfying:
\begin{equation}
\label{M3.16}
T_{zz}=\frac{1}{\gamma^2}\left\{ f; z\right \}~.
\end{equation}  

So-far we have excluded metric singularities, whose presence either in the interior or on the boundary, we will now permit. Then, depending on the conjugacy classes of the monodromy of $A$ and $B$ and of the nature of the metric singularity, there are three classes of local solutions: elliptic, parabolic and hyperbolic. For the present paper the elliptic case is the important one.
\begin{enumerate}
  \item Elliptic : the solution has a curvature singularity, and so (e.g. here with a curvature source at $z=0$ and for $a\in\R$) the Liouville equation reads 
 \begin{equation}
\label{M3.37}
\frac{1}{4\pi}\Delta\phi-\frac{\mu}{8\pi\gamma}\, e^{\gamma\phi}+\frac{1-a}{\gamma}\,\delta^{(2)}(z)=0~,
\end{equation}
Geometrically this is the situation corresponding to conical singularities / corners (orbifolds).
  \item Parabolic : corresponds to punctured Riemann surfaces / surfaces with infinite cusps.
  \item Hyperbolic :  corresponds to a constant negative curvature metric on the annulus, i.e. ``plumbing fixture metric".
\end{enumerate}

There is a simple way of producing conical singularities, more precisely corners. Let us consider some polygon (details will follow in later parts of the paper). Then by the inverse of the standard Schwarz-Christoffel mapping we can biholomorphically map the polygon onto the upper half-plane, such that the vertices get mapped onto the real axis. By pulling-back  the reference  metric on the polygon, we get a metric with corner singularities on $\mathbb{R}$.   
\subsection{Conical singularities and the space of Polygons}

Let us consider a bordered Riemann surface $M$.  A (real) divisor on $M$ is the formal sum
\begin{displaymath}
{\boldsymbol \beta}=\sum_i\beta_i p_i
\end{displaymath}
where the $p_i\in M$ are points and $\beta_i\in\R$.  The discrete set $\{p_i\}$ is the support of ${\boldsymbol\beta}$ and the number $|{\boldsymbol \beta}|:=\sum_i\beta_i$ is the degree of the divisor. We shall have the following conditions on the divisor:
\begin{equation}
\label{ }
\beta_i>-1\quad\text{if}\;\; p_i\notin\partial M\quad\text{and}\quad\beta_i>-\frac{1}{2}\quad\text{if}\;\; p_i\in\partial M~.
\end{equation}
We shall call a simply connected domain $D$ with a divisor $\{(p_1,\beta_1),\dots,(p_n,\beta_n)\}$, a weighted domain. 

Then a conformal metric $ds^2$ on $M$ represents the divisor $\boldsymbol \beta$ if $ds^2$ is a $C^2$-Riemannian metric  on $M\setminus\supp({\boldsymbol \beta})$ such that if $z_i$ is a local coordinate on a neighbourhood $U_i$ of $p_i$, then there exists a continuous function $u:U_i\rightarrow\R$, of class $C^2$ on $U_i\setminus\{p_i\}$, such that on $U_i$:
\begin{equation}
\begin{cases}
ds^2 = e^{2u}|z_i-a_i|^{2\beta_i}|dz_i|^2 &\text{if}\;\; p_i\notin\partial M, \\
ds^2 = e^{2u}|z_i-a_i|^{4\beta_i}|dz_i|^2 &\text{if}\;\; p_i\in\partial M,
\end{cases}
\end{equation}
where $a_i=z_i(p_i)$.  

The point $p_i$ is called a {\em conical singularity of angle} $\theta_i=2\pi(\beta_1+1)$ if $p_i\notin\partial M$ and a {\em corner} of angle $\varphi_i=2\pi(\beta_i+\frac{1}{2})$ or of exterior angle $-2\pi\beta_i$ if $p_i\in\partial M$. In both cases, we shall say that $ds^2$ has a singularity of order $\beta_i$ at $p_i$. Riemann surfaces with conical singularities are generalised Riemann surfaces (GRS) but the natural morphisms for GRS's are still conformal mappings, which topologically are covering maps in the sense of 2 dimensional orbifold theory. The problem of prescribing curvature for such surfaces, has been studied in~\cite{Troyanov:1991}. 

Now, the Gauss-Bonnet formula does still hold in this situation, and it states for a GRS with divisor $(M, {\boldsymbol \beta})$ that:
\begin{equation}
\label{GB}
\frac{1}{2\pi}\int_{M}  R\, \text{dvol}+\frac{1}{2\pi}\int_{\partial M} k\, |dz|=\chi(M, {\boldsymbol \beta})~,
\end{equation}
where the Euler characteristic of $(M, {\boldsymbol \beta})$ is defined by
$\chi(M, {\boldsymbol \beta})\equiv\chi(M)+|{\boldsymbol \beta}|$ with $\chi(M)$ the topological Euler characteristic of $M$. In a physical context this corresponds to some conservation law, e.g. charge conservation. 

As the above discussion shows, polygons are natural objects to consider, if one wants to deal with conical singularities (corners). 

Therefore let us briefly mention some facts about the set ${\mathcal P}_n$ of all polygons with $n\geq3$ distinguished vertices in the complex plane $\mathbb{C}$, whose sides have non-negative length. We shall allow for all possible degenerations of the polygons, with the exception of the degeneration to a single point.  On the complex plane $\C$ we shall consider the usual Euclidean metric $|dz|^2$.

Two polygons are called equivalent if there is an orientation preserving similarity of the complex plane, which maps vertices of one polygon to those of the other one. We know, that these conformal mappings are given by a global linear transformation of the form:
\begin{displaymath}
f(z):=az+b,\quad\text{where}\quad (a,b)\in\C^*\ltimes\C~.
\end{displaymath}
It is also well known that the infinitesimal generators of these transformations are the differential operators:
$\partial_z$ (translations) and $z\partial_z$ (dilatations and rotations) 

If we denote the edges of the $n$-gon $D$ by $e_1,\dots, e_n$ and its vertices by $v_1,\dots, v_n$, then we have the basic relation from vector calculus $\vec{e}_j=v_{j+1}-v_{j}$. So the space ${\mathcal P}_n$ is canonically isomorphic to the complex projective space $\C{\mathbb{P}}^{n-2}$. To see this, just consider the hyperplane $H\subset\C^n$, defined by:
\begin{displaymath}
H:=\{(e_1,\dots, e_n)\in\C^n~:~ e_1+\dots +e_n=0 \}~.
\end{displaymath}
Therefore, the space ${\mathcal P}_n$ is connected and is naturally endowed with the Fubini-Study metric. Hence, every polygon $D\in{\mathcal P}_n$ can be continuously deformed to any polygon $P\in{\mathcal P}_n$.

\subsection{The Correlator Toolbox}
As the present article deals with stochastic processes of $\text{SLE}(\kappa,\rho)$ type, which are basically stochastic multi-particle systems, where the individuals are confined to the boundary of some surface, originally the real line, in a physical context we have  to consider, what kind of particles we are dealing with, e.g.  what quantum numbers they posses.  Whereas in probability theory one might just see a random dynamical system, the physicist perceives them (in the SLE context) rather as objects (quanta) belonging to an underlying field theory. 

\begin{figure}[ht]
\begin{center}
\label{DBrane}
\includegraphics[scale=0.6]{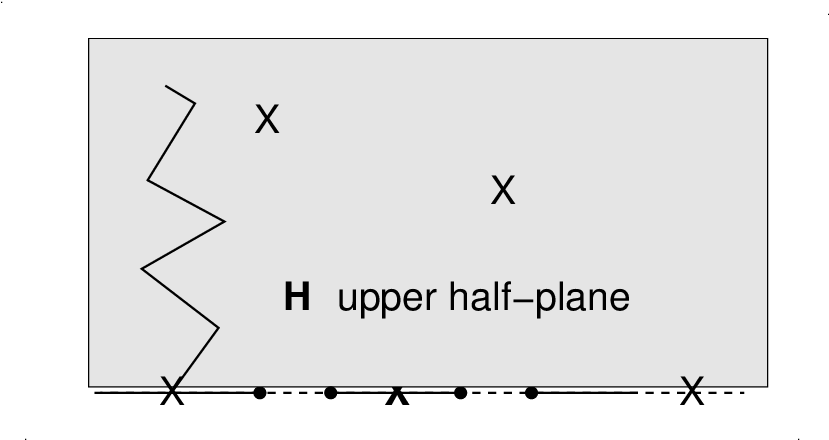}
\caption{The open-string world sheet with changing Neumann and Dirichlet boundary conditions and with bulk and boundary field insertions.}
\label{DB}
\end{center}
\end{figure}

As our previous discussion showed, there are several quantum field theories which are related to the realm of conformal symmetry. They are derived from some classical action and quantized via the path-integral formalism, which in turn, and very loosely speaking, corresponds to the It{\^{o}} Integral. Below, the situation is depicted schematically, where $\phi$ stands for the relevant fields (or just a dummy variable):
\begin{displaymath}
\underbrace{\underbrace{\underbrace{(\nabla\phi)^2}_{\text{BCFT}}\quad QR\phi}_{\text{CGF}}\quad e^{-\gamma\phi}}_{\text{BLFT}}\quad\text{+ boundary terms + boundary conditions}+\text{ghosts}
\end{displaymath}
Further, we also saw, that the quantum field theories discussed, have various characteristic operators of type, e.g.:
\begin{itemize}
  \item bulk / boundary vertex operators
  \item boundary condition changing operators (Cardy type) 
  \item twist fields: Dirichlet$\leftrightarrow$Neumann (``dual resonance theory" type)
\end{itemize}
but also 
\begin{itemize}
  \item bulk and boundary curvature sources of elliptic or parabolic type
\end{itemize}
As the fundamental quantities in a quantum field theory are given by correlators, e.g. $Z$ the partition function, we could try to 
build out of the above operators an arbitrary ``correlator burger", 
\begin{displaymath}
\langle\cdots\text{plug in operators}\cdots\rangle
\end{displaymath}
In the simplest case, it could be derived from the situation in Fig.~\ref{DB}, where we have on the real axis different boundary conditions, Dirichlet and Neumann or discontinuously changing boundary conditions of the same type, curvature sources in the bulk or on the boundary.  

But, as it is known, conformal field theories~\cite{BPZ} are controlled by tight algebraic structures, e.g. the central charge and the dimensions of the fields are dictated by representation theory of, e.g. the Virasoro algebra.  Therefore, not all of the possible insertions would lead to well defined expressions.   

Further, as it was already known in the early days of CFT, that when a conformal field theory is considered on a Riemann surface, the moduli will enter explicitly~\cite{EO1989}. Examples, of ``natural" Riemann surfaces with non-trivial moduli are the unit disc with $10^{26}$ marked points, the annulus or an $n$-connected domain, to mention some of the planar ones.

So, as an illustration, for the Ising model in highest-weight representation with $c=1/2$ there exists a degeneracy at level two with highest-weight state $h=1/2$, i.e. 
\begin{equation}
\label{EO2}
[L_{-2}-\frac{3}{4}(L_{-1})^2]|h=1/2\rangle=0,
\end{equation}
and therefore this state must decouple from the other states in an unitary irreducible representation. This leads, as a particularity of CFT, to a differential equation for correlation functions, i.e.  
there is a correspondence of null states in the representation space and linear differential operators whose order is given by the grade of the null state. 

In the case of a non-trivial Riemann surface, the Ward identities acquire an additional term, which describes the dependence of correlation functions on the moduli of the underlying surface (as a function of the topological invariants, e.g. genus, boundary components, marked points). 

Although the local algebraic structure~(\ref{EO2}) remains the same on the  Riemann surfaces, the identity~(\ref{EO2}) becomes an operator identity,
\begin{equation}
\label{EO6}
(L_{-2}(w)-{\frac{3}{4}}\left(L_{-1}(w))^2\right)\phi_{h=1/2}(w)=0.
\end{equation}
where $\phi$ is a conformal field of dimension $h$.

Now, in the SLE / CFT context, we are interested in certain choices of operators within a correlator, that geometrically seen, create at least one simple curve, e.g. a domain wall, and in possible conformally invariant probability measures supported by these curves. Of course, the measures will depend also on the presence of the other fields. At this point one may consult~\cite{Cardy2004}.

By the Loewner mapping, i.e. by cutting the surface, we can evolve the correlator, to obtain the relevant driving process, and hence by pull-back the probability measure itself. However the stochastic driving process now lives on the appropriate moduli space, and is not a simple one-dimensional Brownian motion, any longer. In case, of conformal invariance and a Markovian-type property the process will be a Markov process, see~\cite{BF:2005a}.  

A look at Fig.~(\ref{DB}), immediately reveals  that the situation can be naturally extended to the case, where we have also insertions of some bulk fields and / or higher loop diagrams.

Therefore, a natural general correlator, which is still describable within the generalised SLE framework, leads to processes of the form
\begin{equation}
\label{krho}
\text{SLE}(\kappa,\vec{\rho}_{\text{boundary}},\vec{\rho}_{\text{bulk}})~.
\end{equation}
The consideration of these, should give new and interesting classes of stochastic processes with associated measures.

\section{An Illustration: Diffusing Polygons}
\begin{figure}[ht]
\begin{center}
\includegraphics[scale=0.5]{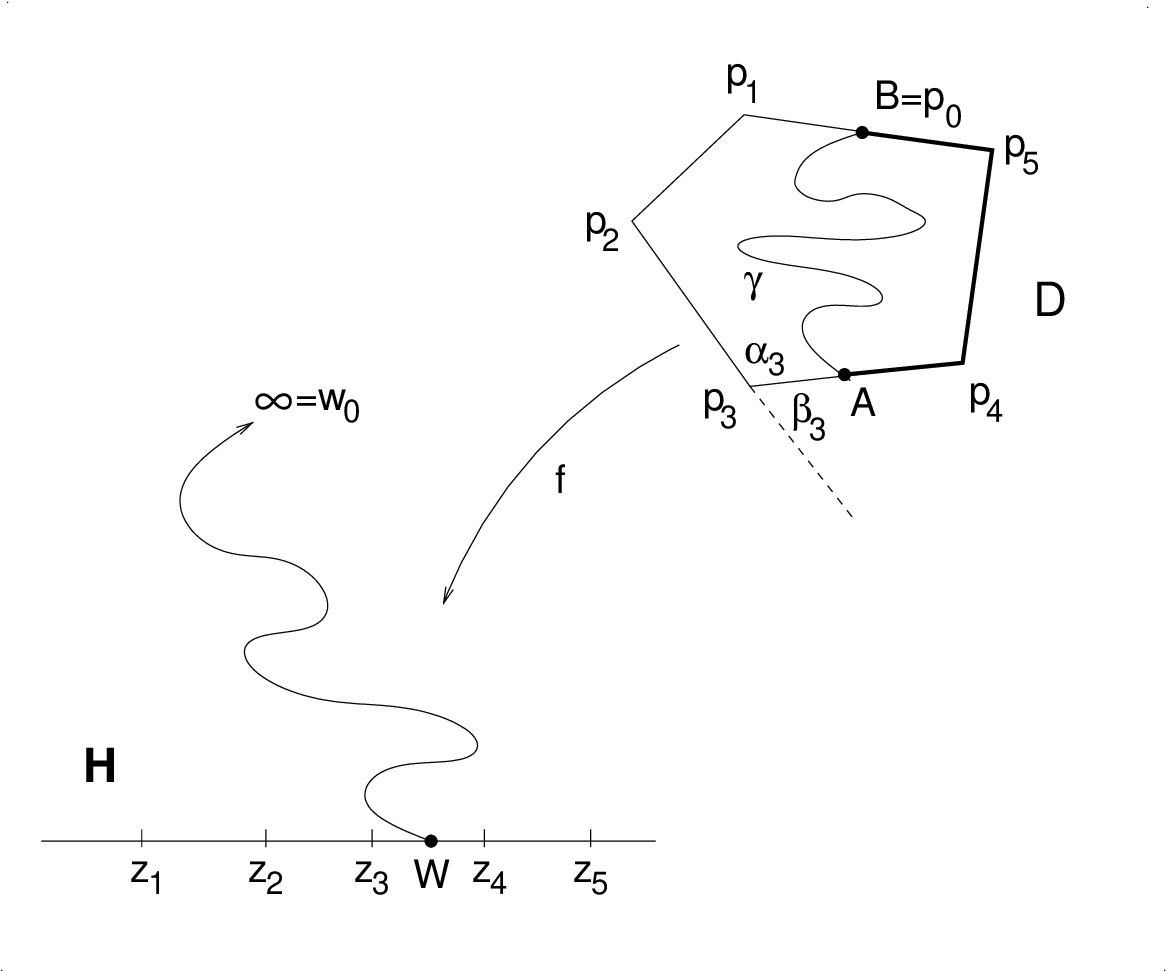}
\caption{A polygon $D$ with changing boundary conditions at $A$ and $B$, domain wall $\gamma$, and the conformal image of the set-up under $f$ on $\overline{\mathbb{H}}$.}
\label{PSLE}
\end{center}
\end{figure}

Let us now illustrate the preceding  discussion with an example. To do so,  let us consider a critical statistical mechanics model defined on a polygon $D$, with changing boundary conditions at points $A$ and $B$, see Fig.~(\ref{PSLE}). As in the usual SLE set-up, this generates a domain wall, that connects the points $A$ and $B$ and in principle is described by a ``sort of chordal SLE".  We will now derive the driving markov process in this situation from purely geometric considerations. 

\subsection{Schwarz-Christoffel Formula.} 
Now let the consecutive vertices be $p_1,\dots p_n$ in positive cyclic order. The angle at $p_k$ is $\alpha_k\pi$, $0<\alpha_k<2$, and the  outer angle is $\beta_k\pi=(1-\alpha_k)\pi$, $-1<\beta_k<1$. We note that 
\begin{equation}\label{E:beta sum}
        \beta_1+\cdots+\beta_n=2~,
\end{equation} 
and that the polygon is convex if and only if all $\beta_k>0$. We will call the pairs $(p_k,\beta_k)$ the corners of the polygon.

Further, let  $f$ be a conformal map from $D$ onto the upper half-plane $\mathbb H$, with $z_k=f(p_k)$ and such that none of the $z_k$ equals $\infty$. Then for $z\in\mathbb H$ define the Schwarz-Christoffel mapping
\begin{equation}\label{E:sc}
        SC(z)=SC\left[\begin{array}{c | c}z_1, \dots, z_n &  z \\
        \beta_1, \dots,  \beta_n  & z^*\end{array}\right]=\int_{z^*}^z\prod_{k=1}^n(z-z_k)^{-\beta_k}\ dz,
\end{equation}
where the powers $(z-z_k)^{-\beta_k}$ denote analytic branches in $\mathbb H$. Note that 
\begin{equation}\label{E:SCprime}
        SC'(z)=\prod_{k=1}^n (z-z_k)^{-\beta_k},
\end{equation}
and
\begin{equation}\label{E:SC}
        \frac{SC''(z)}{SC'(z)}=-\sum_{k=1}^n\frac{\beta_k}{z-z_k}. 
\end{equation}
Then it is well known that for some constants $a,b\in \mathbb C$, $f^{-1}= a SC+b$, 
see \cite{ahlfors:1966}. This result extends to the case when the polygon is allowed to have slits, i.e $\beta_k=-1$ for some $k$. Slits are counted as double edges of the boundary polygon, traversed in positive cyclic order. A vertex, when considered as a boundary point, may then occur multiple times, corresponding to different prime ends. Henceforth, a corner $(p_k,\beta_k)$ will always be a pair consisting of a prime end $p_k$ located at a vertex together with the exterior angle $\beta_k$ associated to the prime end $p_k$. The formula \eqref{E:sc} then remains unchanged if $z_k=f(p_k)$.
In case $f(p_k)=\infty$ for one $k$, then \eqref{E:sc} needs to be adjusted by simply dropping the factor with exponent $-\beta_k$. 
Finally, formula \eqref{E:sc} continues to hold if $D$ is unbounded or one (or several) corners are at $\infty$, provided that the angles at $\infty$ are appropriately defined, see \cite{henrici:1974}. The interior angle at $\infty$ is chosen in $[-2\pi,0]$. So, polygon will refer to these ``generalised" polygons. 

In the case of polygons, we still have, that if  $D$ is a polygon with corners $(p_1,\beta_1)$ to $(p_n,\beta_n)$ in positive cyclical order, and $\gamma$ is a Jordan arc contained in $D$ except for one endpoint which lies on the interior of a side $S$ of $D$, then there is a conformal map $f$ from $D\backslash\gamma$ onto a polygon $D'$ such that $(f(p_1),\beta_1),\dots, (f(p_n),\beta_n)$ are the corners of $D'$ in positive cyclical order. If we require $f(S\cup\gamma)\subset[0,1]$,  then $f$ is unique.

\section{SLE$(\kappa,\rho)$ and Polygon motion}
 
Let $(p_1,\beta_1),\dots,(p_n,\beta_n)$ be the corners of a polygon $D$. Denote $f$ a conformal map from $D$ onto the upper half-plane and set $z_k=f(p_k)$, $1\le k\le n$. We assume the points $z_k$ are all finite and distinct from $0$. For $\kappa>0$ set
\begin{equation}\label{E:rho-mu}
        \rho_k=\frac{\kappa}{2}\beta_k,\quad k=1,\dots, n.
\end{equation}
In particular, $-\kappa/2\le\rho_k\le\kappa/2$.
Suppose that $(W_t, Z^1_t,\dots Z_t^n)$ is a solution to \eqref{E:rho-diff}. 
 For $z$ in the upper half-plane, set
\[
        SC_t(z)=SC\left[\begin{array}{c | c}Z_t^1, \dots, Z_t^n & z \\
        \beta_1, \dots,  \beta_n & 0\end{array}\right].
\]
Then $z\mapsto SC_t(z)$ extends continuously to the real axis with the points $Z_t^k$ removed and is differentiable there as a function of $t$. In particular, if $W_s \neq Z_s^1,\dots, Z_s^n$ for $s\in[0,t]$, then we may define
\begin{equation}\label{E:def f}
        f_t(z)=SC_t(z)-\int_0^t(\partial_s SC_s)(W_s)\ ds.
\end{equation} 
Note that $f_t$ maps $\mathbb{H}$ onto a polygon while the function $f$ as previously introduced or in Fig.~(\ref{PSLE}), maps a polygon onto the upper half-plane, i.e. is the inverse Schwarz-Christoffel mapping. Define the stopping time $\sigma$ by
\[
\sigma=\sup\,\{t : W_s,\, Z^1_s,\dots, Z_s^n\;\text{are all distinct for $0\leq s\leq t$}\,\}~.
\]

Then the process $U_t\equiv f_t(W_t)$ is a martingale for $t<\sigma$ and  if 
\[
        A_t\equiv\kappa\int_0^t\left(SC'_s(W_s)\right)^2\ ds
\]
and $\tau(t)$ is defined by $A_{\tau(t)}=t$, then $t\mapsto U_{\tau(t)}$ is a standard Brownian motion. We further note, that the motion of the corners of the polygon $f_t(\mathbb H)$ is
differentiable. 

It is important to point out, that if we begin with an arbitrary SLE($\kappa,\rho$), i.e. with a choice of points $z_1,\dots, z_2$ and weights $\rho_1,\dots,\rho_n$, then the results of this section will still hold. However, in this case the Schwarz-Christoffel mapping $SC$ is no longer to be schlicht, but it still maps the intervals $[z_k, z_{k+1}]$ onto straight line segments. But, by considering the associated Riemann surface to the analytic function $SC$, we can still interpret the image $SC(\mathbb{H})$ as a Riemannian domain. So, in the case of SLE($2,(-1,-1))$, up to a normalisation, this corresponds to the map $z^3-3z$, which can be understood in terms of a $3$-fold cover, cf.~\cite{ahlfors:1966}

Set now $D_t=f_t(\mathbb H)$, and denote 
\[
(q,u)\in D_t\times\partial D_t\mapsto k_{D_t}(q,u)
\]
the Poisson kernel of $D_t$. If $p\in\partial D_t$, denote $\partial_2 H_{D_t, p}(q,u)$ the analytic function in $q$ whose real part is $\partial_2 k_{D_t}(q,u)$ and which satisfies 
\[
\lim_{q\to p}\partial_2 H_{D_t,p}(q,u)=0~.
\]

Then the main statement is
\begin{theorem}[Loewner evolution in polygons]
Denote $K_t$ the hull of an $\text{SLE}_{\kappa}(\rho)$ in the upper half-plane
and $g_t:\mathbb H\backslash K_t\to\mathbb H$ the normalised uniformising
map. Then $h_t\equiv f_t\circ g_t\circ f_0^{-1}:D\backslash
f_0(K_t)\to D_t$ satisfies
\begin{equation}\label{E:PL}
\partial_t\ln h_t'(z)=f_t(W_t)^2\partial_2 H_{D_t, f_t(\infty)}(h_t(z), f_t(W_t))~.
\end{equation}
\end{theorem}

\section{SLE coupled to Gravity}
As it is know, a stochastic processes is linked to some operator, which intrinsically depends upon a metric. The operator which determines what Brownian motion should be, is the Laplace-Beltrami operator. In local coordinates, applied to a function $\varphi$, it reads as 
\[
\Delta\varphi=\frac{1}{\sqrt g}\frac{\partial}{\partial x^j}\left(\sqrt{g}\,g^{ij}\frac{\partial \varphi}{\partial x^i}\right)~.
\]

So, instead of mapping $\text{SLE}(\kappa,\rho)$ into polygons we can also stay in
the upper half-plane and couple it to a fluctuating background metric. Indeed, $f_t:\mathbb H\to D_t$ is
an immersion. If we endow $D_t$ with the Euclidean metric, then the pull-back metric
via $f_t$ on $\mathbb H$ is 
\[
g_{ij}=\delta_{ij}|f_t'(z)|^2,\quad i,j=1,2,
\]
where the indices 1 and 2 refer to the real and imaginary coordinate,
respectively. If $\Gamma=(\Gamma_{jk}^i)$ denotes the Levi-Civita connection
for this metric, then the (2-dimensional) Brownian motion $\tilde{W}$ for the
metric $(g_{ij})$ solves the stochastic differential equation
\[
d\tilde{W}_s^i=\sigma_j^i(\tilde{W}_s)\ dB_s^j-\frac{1}{2}
g^{kl}(\tilde{W}_s)\Gamma_{kl}^i(\tilde{W}_s)\ ds,
\]
see \cite{hsu:2002}. Here $g^{-1}=(g^{kl})$ is the inverse coefficient matrix
of $g$ and $\sigma$ is a square root of $g^{-1}$ (i.e. $\sigma
\sigma^T=g^{-1}$), and we apply the Einstein summation convention. For our particular metric $g$ we find
\[
\Gamma_{11}^1=\Gamma_{22}^1=-\Re\left(\frac{f_t''}{f_t'}\right),
\]
see \cite{carmo:1992}. The boundary $\mathbb R=\partial\mathbb H$ is a 
one-dimensional submanifold of $\overline{\mathbb H}$. The metric $g$ on
$\mathbb H$ thus induces the metric $(f_t'(x))^2\ dx^2$ on $\mathbb R$. A
(one-dimensional) Brownian motion $W$ relative to this metric solves the
stochastic differential equation
\begin{equation}\label{E:metricBM}
dW_s=\frac{dB_s}{f_t'(W_s)}-\frac{1}{2(f_t'(W_s))^2}\sum_{j=1}^n\frac{\beta_j}{W_s-Z^k_t}\
ds.
\end{equation}
We now couple the metric to the Brownian motion $W$ via
\begin{equation}\label{E:couple}
dZ_t^k=\frac{2}{\kappa(f_t'(W_t))^2(Z^k_t-W_t)}\ dt,\quad k=1,\dots, n.
\end{equation}
Then, after a timechange, \eqref{E:metricBM} and \eqref{E:couple} become the
$\text{SLE}(\kappa,\rho)$-system \eqref{E:rho-diff} with the convention
$\rho_j=\kappa \beta_j/2$.

\subsubsection*{Acknowledgements}
R.~F. would like to thank Shoibal Chakravarty for the questions asked and the discussions.  Gast\'on Giribert he thanks for helpful explanations. John Cardy he thanks  for discussing the notes and for general discussions.

{The research of the first author was supported by 	NSA grant H98230-04-1-0039.}\\
{The research of the second author was supported by a grant of the Max-Planck-Gesellschaft.}

\end{document}